\documentclass[10pt,conference]{IEEEtran}

\usepackage{mathbf-abbrevs}

\newcommand{\reals}{{\mathbb R}}
\newcommand{\ints}{{\mathbb Z}}

\newcommand{\vor}{\operatorname{Vor}}

\newcommand{\term}{\emph}

\newcommand{\abs}[1]{{\left| #1 \right|}}

\newcommand{\dotprod}[2]{ #1 \cdot #2}

\usepackage[square,comma,numbers,sort&compress]{natbib}

\usepackage{units}
		
\usepackage{booktabs}
		
\usepackage{ifpdf}
\ifpdf
  \usepackage[pdftex]{graphicx}
\else
	\usepackage{graphicx}
\fi

\usepackage{amsmath,amsfonts,amssymb, amsthm, bm} 

\usepackage[vlined, linesnumbered]{algorithm2e}
	
\usepackage{mathrsfs}

\newtheorem{theorem}{Theorem}

\newtheorem{corollary}{Corollary}

\newtheorem{definition}{Definition}

\begin{document}

\title{Finding short vectors in a lattice of Voronoi's first kind}

\author{\IEEEauthorblockN{Robby~McKilliam and Alex~Grant}
\IEEEauthorblockA{Institute for Telecommunications Research, The University of South Australia, SA, 5095}
}

\markboth{Finding short vectors in a lattice of Voronoi's first kind}{\today}

\maketitle

\begin{abstract}
We show that for those lattices of Voronoi's first kind, a vector of shortest nonzero Euclidean length can computed in polynomial time by computing a minimum cut in a graph.
\end{abstract}


\section{Introduction}\label{sec:introduction}

A $n$-dimensional \term{lattice}, $\Lambda$, is a discrete set of vectors from $\reals^m$, $m \geq n$, formed by the integer linear combinations of a set of linearly independent basis vectors $b_1, \dots, b_n$ from $\reals^m$.  That is, $\Lambda$, consists of all those vectors, or \emph{lattice points}, $x \in \reals^m$ satisfying
\[
  x = b_1 u_1 + b_2u_2 + \dots + b_n u_n \qquad u_1, \dots u_n \in \ints.
 \] 


An interesting question about a lattice is: `What is the shortest distance between any two lattice points?'  Because the origin is a lattice point this question can be equivalently stated as: `What is the length of the shortest lattice point not equal to the origin?'   Those points in the lattice with shortest length are called \emph{short vectors}.  The problem of discovering a lattice point of shortest length is called the \emph{shortest vector problem} and has applications to, for example, cryptography~\citep{Micciancio_lattice_based_post_quantum_crypto}.  In general, i.e. for arbitrary lattices, the shortest vector problem is NP-hard~\cite{micciancio_hardness_2001,AjtaiShortestVecProbNPHard1998}\footnote{Specifically, the shortest vector problem is known to be hard for NP under what are called \emph{reverse unfaithful random} reductions~\cite{micciancio_hardness_2001}.}.  However, for some lattices, the problem is easy to solve.  For example, short vectors are relatively easy to determine in highly regular lattices, such as the root lattices $A_n$ and $D_n$, their dual lattices $A_n^*$ and $D_n^*$, and the integer lattice $\ints^{n}$~\cite[Chap. 4]{SPLAG}.

In this paper we consider a particular class of lattices, those of \emph{Voronoi's first kind}~\cite{ConwaySloane1992_voronoi_lattice_3d_obtuse_superbases,Valentin2003_coverings_tilings_low_dimension,Voronoi1908_main_paper}.  We show that a short vector in a lattice of Voronoi's first kind can be computed in polynomial time by computing a minimum cut in a graph.  If the lattice has dimension $n$, this requires $O(n^3)$ operations to deterministically compute a short vector~\cite{Stoer_simple_min_cut_1997}, or $O(n^2(\log n)^3)$ operations to compute a short vector with high probability~\cite{Karger_mincut_ranomised_1996}. 

The paper is structured as follows.  Section~\ref{sec:voron-cells-relev} defines the short vectors of a lattice along with some related entities, the \emph{Voronoi cell} and the \emph{relevant vectors}.  Section~\ref{sec:latt-voron-first} defines lattices of Voronoi's first kind and shows how their short vectors can be found by solving a constrained quadratic program in binary $\{0,1\}$ variables.  Section~\ref{sec:quadratic-0-1} describes how this constrained program can be mapped into that of computing a minimum cut in a weighted graph.
Some examples are given in Section~\ref{sec:some-examples}.

\section{Voronoi cells, relevant vectors, and short vectors}\label{sec:voron-cells-relev}
\newcommand{\calR}{\mathcal{R}}
The (open) \term{Voronoi cell}, denoted $\vor(\Lambda)$, of a lattice $\Lambda$ in $\reals^n$ is the subset of $\reals^n$ containing all points nearer, with respect to a given norm, the lattice point at the origin than any other lattice point. The Voronoi cell is an $n$-dimensional convex polytope that is symmetric about the origin. Here we will always assume the Euclidean norm (or 2-norm), so $\vor(\Lambda)$ contains those points nearest in Euclidean distance to the origin. 

 
Equivalently the Voronoi cell can be defined as the intersection of the half spaces 
\[
H_{v} = \{x \in \reals^n \mid \dotprod{x}{v} < \tfrac{1}{2}\dotprod{v}{v} \}
\]
for all $v \in \Lambda \backslash  \{\zerobf\}$, i.e. all lattice points $v \in \Lambda$ not equal to the origin $\zerobf$.  Here $\dotprod{x}{v}$ is the inner product between vectors $x$ and $v$.  It is not necessary to consider \emph{all} $v \in \Lambda \backslash  \{\zerobf\}$ to define the Voronoi cell.  The minimal set of lattice vectors $\calR$ such that $\vor(\Lambda) = \cap_{v\in\calR}{H_{v}}$ is called the set of \term{Voronoi relevant vectors} or simply \term{relevant vectors}~\citep{Voronoi1908_main_paper}. 

\begin{definition} The relevant vectors $v \in \Lambda$ are those for which  
\[
\dotprod{v}{x} < \dotprod{x}{x}
\]
for all $x \in \Lambda$ with $x \neq v$ and $x \neq \zerobf$.\footnote{These are the `strict' relevant vectors according to Conway and Sloane~\cite{ConwaySloane1992_voronoi_lattice_3d_obtuse_superbases}.  If the inequality $\dotprod{v}{x} < \dotprod{x}{x}$ is replaced by $\dotprod{v}{x} \leq \dotprod{x}{x}$ then this would also include the `lax' relevant vectors.  The short vectors are always strict so we only have use of the strict relevant vectors here.}
\end{definition}

The \emph{short} (or \emph{minimal}) vectors in a lattice are all those lattice points of minimum nonzero Euclidean length, i.e all those of squared length
\[
\min_{x\in \Lambda \backslash \{ \zerobf \} } \| x \|^2.
\]  
Plainly, every short vector is also a relevant vector, because, if a lattice point $s$ is not relevant there exists a lattice point $x \neq s$ such that $\dotprod{x}{s} < \dotprod{s}{s}$, implying that $\|x\|^2 < \|s\|^2$, so $s$ is not a short vector.


\section{Lattices of Voronoi's first kind}\label{sec:latt-voron-first}

An $n$-dimensional lattice $\Lambda$ is said to be of \emph{Voronoi's first kind} if it has what is called an \emph{obtuse superbase}~\cite{ConwaySloane1992_voronoi_lattice_3d_obtuse_superbases}.  That is, there exists a set of $n+1$ vectors $b_1,\dots,b_{n+1}$ such that $b_1,\dots,b_n$ are a basis for $\Lambda$,
\begin{equation}\label{eq:superbasecond}
b_1 + b_2 \dots + b_{n+1} = 0
\end{equation}
(the \emph{superbase} condition), and the inner products satisfy,
\[
q_{ij} = b_i \cdot b_j \leq 0, \qquad \text{for} \qquad i,j = 1,\dots,n+1, i \neq j
\]
(the \emph{obtuse} condition).  The $q_{ij}$ are called the \emph{Selling parameters}.  It is known that all lattices in dimensions less than $4$ are of Voronoi's first kind~\cite{ConwaySloane1992_voronoi_lattice_3d_obtuse_superbases}.  An interesting property of lattices of Voronoi's first kind is that their relevant vectors have a straightforward description.

\begin{theorem} (Conway and Sloane~\cite[Theorem~3]{ConwaySloane1992_voronoi_lattice_3d_obtuse_superbases})
The relevant vectors of $\Lambda$ are of the form,
\[
\sum_{i \in I} b_i
\]
where $I$ is a strict subset of $\{1, 2, \dots, n+1\}$ that is not empty, i.e. $I \subset \{1, 2, \dots, n+1\}$ and $I \neq \emptyset$.
\end{theorem}


Because each short vector is also a relevant vector we have the following corollary.

\begin{corollary}\label{thm:minvectorisrel}
The short vectors in $\Lambda$ are of the form $\sum_{i \in I} b_i$ where $I \subset \{1, 2, \dots, n+1\}$ and $I \neq \emptyset$.
\end{corollary}

Given this corollary a na\"{i}ve way to compute a short vector is to compute the squared length $\| \sum_{i \in I} b_i \|^2$ for all of the $2^{n+1} - 2$ possible $I$ and return a lattice point with minimum squared length.  This procedure requires a number of operations that grows exponentially with the dimension $n$.  We can improve this using a minimum cut algorithm.  
To facilitate this consider the quadratic form
\begin{equation}\label{eq:quadformnp}
Q(u) = \| \sum_{i=1}^{n+1} b_iu_i \|^2 =  \sum_{i=1}^{n+1}\sum_{j=1}^{n+1} q_{ij}u_i u_j.
\end{equation}
A short vector is a minimiser of this form under the constraint that the $u_i \in \{0,1\}$ and at least one element of $u$ is equal to $1$ and at least one element of $u$ is equal to $0$.  The next section will show how this constrained quadratic minimisation problem can be solved by computing a minimum cut in a graph.  This technique has appeared previously~\cite{Picard_min_cuts_1974,Sankaran_solving_CDMA_mincut_1998,Ulukus_cdma_mincut_1998} but we include the derivation here so that this paper is self contained.  

\section{Quadratic $\{0,1\}$ programs and minimum cuts in graphs}\label{sec:quadratic-0-1}

Let $G = \{ V, E \}$ be an undirected graph with $n+1$ vertices $v_1, \dots, v_{n+1}$ contained in the set $V$ and edges $e_{ij} \in E$ connecting vertex $v_i$ to vertex $v_j$.  To each edge we assign a \emph{weight} (or \emph{capacity}) $w_{ij} \in \reals$.  The graph is undirected so the weights are symmetric, i.e. $w_{ij} = w_{ji}$.  A \emph{cut} in the graph $G$ is a nonempty subset $C \subset V$ of vertices with its (also nonempty) complement $\bar{C} \subset V$.  That is, a cut is the pair $(C, \bar{C})$ such that both $C$ and $\bar{C}$ are not empty, $C \cap \bar{C} = \emptyset$ and $C \cup \bar{C} = V$.

The weight of a cut is
\[
W(C,\bar{C}) = \sum_{i \in I} \sum_{j \in J} w_{ij}, 
\]
where $I = \{ i \mid v_i \in C\}$ and $J = \{j \mid v_j \in \bar{C}\}$.  That is, $W(C,\bar{C})$ is the sum of the weights on the edges crossing from the vertices in $C$ to the vertices in $\bar{C}$.  If the graph is allowed to contain loops, i.e. edges from a vertex to itself, then the weight of these edges $w_{ii}$ have no effect on the weight of any cut.  We may choose any values for the $w_{ii}$ without affecting $W(C,\bar{C})$.  

The \emph{minimum cut} is the $C$ and $\bar{C}$ that minimise the weight $W(C,\bar{C})$.  Let $\abs{V} = n+1$ and $\abs{E} \leq \tfrac{1}{2}(n+1)n$ denote the number of vertices and the number of edges in the graph $G$.  If all of the edge weights $w_{ij}$ for $i \neq j$ are nonnegative, a minimum cut can be computed deterministically in order 
\[
O(\abs{V}\abs{E} + \abs{V}^2\log{\abs{V}}) \in O(n^3)
\]
 arithmetic operations using the algorithm of Stoer and Wagner~\cite{Stoer_simple_min_cut_1997} and with high probability in 
\[
O(\abs{V}^2 (\log\abs{V})^3) = O(n^2 (\log(n))^3)
\]
operations using the randomised algorithm of Karger and Stien~\cite{Karger_mincut_ranomised_1996}.

We now show how $W(C,\bar{C})$ can be represented as a quadratic form.  Define the vector $u$ of length $n+1$ so that 
\[
u_i = \begin{cases}
1, & i \in C \\
0, & i \in \bar{C}.
\end{cases}
\]
Then
\[
u_i(1 - u_j) = \begin{cases}
1, & i \in C, j \in \bar{C} \\
0, & \text{otherwise}.
\end{cases}
\]
The weight can now be written as
\begin{align*}
W(C,\bar{C}) = \sum_{i \in C} \sum_{j \in \bar{C}} w_{ij} = \sum_{i =1}^{n+1} \sum_{j =1}^{n+1} w_{ij} u_i (1 - u_j) = F(u),
\end{align*}
say.  Finding a minimum cut is equivalent to finding the binary $\{0,1\}$ vector $u$ that minimises $F(u)$ under the constraint that at least one element in $u$ is equal to $1$ and at least one element in $u$ is equal to $0$.  This constraint corresponds to the requirement that both $C$ and $\bar{C}$ are nonempty.

Expanding,
\begin{align*}
F(u) &=  \sum_{i=1}^{n+1} \sum_{j =1}^{n+1} w_{ij}u_i - \sum_{i=1}^{n+1} \sum_{j =1}^{n+1} w_{ij} u_iu_j \\
&= \sum_{i=1}^{n+1} k_i u_i - \sum_{i=1}^{n+1} \sum_{j =1}^{n+1} w_{ij} u_iu_j,
\end{align*}
where $k_i = \sum_{j =1}^{n+1} w_{ij}$.  Observe the equivalence of $F(u)$ and $Q(u)$ from~(\ref{eq:quadformnp}) when the weights are assigned according to,
\[
q_{ij} = - w_{ij} \qquad \text{for} \qquad \,\, i,j = 1,\dots,n+1.
\]
Note that with these weights $k_i = -\sum_{j =0}^{n+2} q_{ij} = 0$ due to the superbase condition~\eqref{eq:superbasecond}.

Because the $q_{ij}$ are nonpositive for $i \neq j$ the weights $w_{ij}$ are nonnegative for all $i \neq j$ with $i,j \in \{1,n+1\}$.  As discussed the value of the weights $w_{ii}$ have no effect on the weight of any cut so setting $q_{ii} = - w_{ii}$ for  $i \in \{1,n+1\}$ is of no consequence.  A vector $u$ that minimises $Q(u)$ can be found by computing a minimum cut in the graph with these weights.  A short vector is then given as $\sum_{i=1}^{n+1} b_iu_i$.  This is our main result, so we restate it as a theorem.

\begin{theorem}
Let $\Lambda$ be a $n$-dimensional lattice of Voronoi's first kind with obtuse superbase $b_1, \dots, b_{n+1}$ and Selling parameters $q_{ij} = \dotprod{b_i}{b_j}$.  Let $G$ be a graph with $n+1$ vertices $v_{1}, \dots, v_{n+1}$ and edges $e_{ij}$ with weight $-q_{ij}$.  Let $(C, \bar{C})$ be a minimum cut in the graph $G$.  A short vector in the lattice $\Lambda$ is given by $\sum_{i \in I} b_i$ where $I = \{ i \mid v_i \in C\}$.  The squared Euclidean length of the short vector is given by the weight of the minimum cut $W(C,\bar{C})$.
\end{theorem}

\section{Some examples}\label{sec:some-examples}

As examples we apply this minimum cut approach to finding a short vector in the root lattice $A_n$ and its dual lattice $A_n^*$~\cite[pp. 108-117]{SPLAG}.  An obtuse superbase for $A_n$ is all the cyclic shifts of the vector $\left[ \begin{array}{rrrrrrr} 1 & -1 & 0 & 0 & \cdots & 0 \end{array}\right]$ from $\reals^{n+1}$.  Set the vectors,
\begin{align*}
b_1 &= \left[ \begin{array}{rrrrrrr} 1 & -1 & 0 & 0 & \cdots & 0 \end{array}\right] \\
b_2 &= \left[ \begin{array}{rrrrrrr} 0 & 1 & -1 & 0 & \cdots & 0 \end{array}\right] \\
&\vdots \\
b_{n+1} &= \left[ \begin{array}{rrrrrrr} -1 & 0 & 0 & \cdots & 0 & 1 \end{array}\right].
\end{align*}
The Selling parameters are 
\[
q_{ij} = \dotprod{b_i}{b_j} = \begin{cases} 2, & i = j \\
-1, & i - j \equiv 1 \bmod{n+1} \\
0, & \text{otherwise}.
\end{cases}
\]
The corresponding weighted graph is the cycle graph with $n+1$ vertices, each edge having weight $1$.  A minimum cut in this graph is to choose $C$ to contain $c$ consecutive vertices modulo $n+1$, where $1 \leq c \leq n$.  That is, choose
\[
C = \{ v_{i}, v_{i+1}, \dots v_{i+c} \}
\]
for any integer $i$ where the indices are considered modulo $n+1$.  The weight of such a cut is $2$.  A short vector in $A_n$ is correspondingly  of the form
\[
\sum_{j = i}^{i+c} b_{i+c} =  e_i - e_{i+c}
\]
where $e_i \in \reals^{n+1}$ denotes a vector of all zeros except the $i$th element which is equal to one.  Again, the indices here are considered modulo $n+1$.  The squared Euclidean length of this short vector is $2$.  It follows that short vectors in $A_n$ are of the form $e_i - e_j$ for $i,j \in \{1,\dots,n+1\}$, $i \neq j$, a well known fact~\cite[pp. 108-117]{SPLAG}.  Figure~\ref{fig:an3} displays the graph corresponding to the lattice $A_3$.

An obtuse superbase for the dual lattice $A_n^*$ is all cyclic shifts of the vector 
\[
\left[ \begin{array}{cccc}  \tfrac{n}{n+1} & -\tfrac{1}{n+1} & \cdots & -\tfrac{1}{n+1} \end{array}\right] \in \reals^{n+1}.
\]
Set the vectors,
\begin{align*}
b_1 &= \left[ \begin{array}{rrrrr} \tfrac{n}{n+1} & -\tfrac{1}{n+1} &  -\tfrac{1}{n+1} & \cdots & -\tfrac{1}{n+1} \end{array}\right] \\
b_2 &= \left[ \begin{array}{rrrrr}  -\tfrac{1}{n+1} & \tfrac{n}{n+1} & -\tfrac{1}{n+1}  & \cdots & -\tfrac{1}{n+1} \end{array}\right] \\
&\vdots \\
b_{n+1} &= \left[ \begin{array}{rrrrr} -\tfrac{1}{n+1} &  -\tfrac{1}{n+1}  & \cdots & -\tfrac{1}{n+1} & \tfrac{n}{n+1} \end{array}\right].
\end{align*}
The Selling parameters are
\[
q_{ij} = \dotprod{b_i}{b_j} = \begin{cases} \frac{n}{n+1}, & i = j \\
\frac{-1}{n+1}, & \text{otherwise}.
\end{cases}
\]
The corresponding graph is the complete graph with $n+1$ vertices and $\tfrac{1}{2}(n+1)n$ edges, each edge having weight $\frac{1}{n+1}$.  As each weight is equal and the graph is complete, a minimum cut is given by placing precisely one vertex into $C$ and the remaining vertices into $\bar{C}$.  The weight of this cut is $\frac{n}{n+1}$.  All the superbase vectors $b_1, \dots, b_{n+1}$ are short vectors of squared Euclidean length $\frac{n}{n+1}$, again a well known fact~\cite[pp. 108-117]{SPLAG}.  Figure~\ref{fig:anstar3} shows the graph corresponding to the lattice $A_3^*$.

\begin{figure}[tp]
	\centering 
		\includegraphics{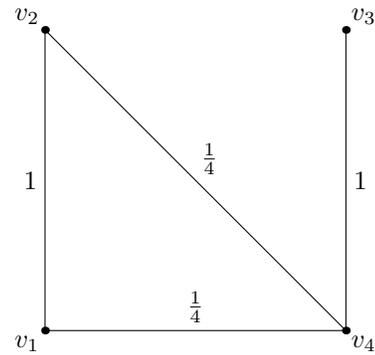}
		\caption{Graph corresponding to the lattice with superbase given by~\eqref{eq:otherlatticesuperbase}.}
		\label{fig:other3lattice}
\end{figure}

In both of the previous examples short vectors could be obtained by picking one of the vectors $b_1, \dots, b_{n+1}$ in the obtuse superbase.  This is not always the case, as our final example will show.  Consider the 3-dimensional lattice with obtuse superbase
\begin{equation}\label{eq:otherlatticesuperbase}
\begin{split}
b_1 &= \left[ \begin{array}{rrr} 1 & -\tfrac{1}{2} & 0 \end{array}\right] \\
b_2 &= \left[ \begin{array}{rrr}  -\tfrac{1}{2} & 1 & 0  \end{array}\right] \\
b_3 &= \left[ \begin{array}{rrr}  0 & 0 & 1   \end{array}\right] \\
b_4 &= \left[ \begin{array}{rrr} -\tfrac{1}{2} &  -\tfrac{1}{2}  & -1 \end{array}\right].
\end{split} 
\end{equation}
The Selling parameters are given in matrix form as
\[
\left[ \begin{array}{cccc} 
q_{11} & q_{12} & q_{13} & q_{14}\\
q_{21} & q_{22} & q_{23} & q_{24} \\
q_{31} & q_{32} & q_{33} & q_{34} \\
q_{41} & q_{42} & q_{43} & q_{44}
 \end{array}\right]
=
\left[ \begin{array}{rrrr} 
\tfrac{5}{4} & -1 & 0 & -\tfrac{1}{4}\\
-1 & \tfrac{5}{4} & 0 & -\tfrac{1}{4} \\
0 & 0 & 1 & -1 \\
-\tfrac{1}{4} & -\tfrac{1}{4} & -1 & \tfrac{3}{2}
 \end{array}\right].
\]
Figure~\ref{fig:other3lattice} displays the corresponding graph.  The minimum cut is given by choosing $C = \{v_1, v_2\}$ and $\bar{C} = \{v_3,v_4\}$.  This corresponds to the short vector
\[
b_1 + b_2 = \left[ \begin{array}{ccc} \tfrac{1}{2} & \tfrac{1}{2} & 0 \end{array}\right]
\]
of squared Euclidean length $\tfrac{1}{2}$.  Note that each of $b_1, b_2, b_3$ and $b_4$ have squared Euclidean length greater than $\tfrac{1}{2}$ and are therefore not short vectors.
\begin{figure}[tp]
	\centering 
		\includegraphics{graphs-1.mps}
		\caption{Graph corresponding to the lattice $A_3$.}
		\label{fig:an3}
\end{figure}

\begin{figure}[tp]
	\centering 
		\includegraphics{graphs-2.mps}
		\caption{Graph corresponding to the lattice $A_3^*$.}
		\label{fig:anstar3}
\end{figure}

\section{Conclusion}

We have described a method from computing a short vector in a lattice of Voronoi's first kind.  This is achieved by mapping the shortest vector problem into that of computing a minimum cut in a weighted graph.  Existing polynomial time algorithms can then be applied~\cite{Stoer_simple_min_cut_1997,Karger_mincut_ranomised_1996}.

\small
\bibliography{bib}

\end{document}